\documentclass[useAMS,usenatbib,usegraphicx]{mn2e}

\usepackage{times}
\usepackage{hyperref}

\newcommand\mnras{MNRAS}
\newcommand\aap{A\&A}
\newcommand\aj{AJ}
\newcommand\actaa{Acta Astron.}
\newcommand\apj{ApJ}
\newcommand\apjl{ApJ}
\newcommand\apjs{ApJSS}

\newcommand\pasp{PASP}
\newcommand\nat{Nature}

\newcommand\qjras{QJRAS}

\title[First results from the {\it Herschel} Gould Belt Survey in Taurus]{First results from the {\it Herschel}\thanks{{\it Herschel} is an ESA space observatory with science instruments provided by European-led Principal Investigator consortia and with important participation from NASA.} Gould Belt Survey in Taurus }

\author[J.M. Kirk et al]{%
{\parbox{\textwidth}{J. M. Kirk$^{1,2}$\thanks{Email: \texttt{jmkirk@uclan.ac.uk}},
D. Ward-Thompson$^{2}$,
P. Palmeirim$^{3}$, 
Ph. Andr\'e$^{3}$,
M. J. Griffin$^{1}$, 
P. J. Hargrave$^{1}$,
V. K\"onyves$^{3,9}$,
J.-P. Bernard$^{4,5}$,
D. J. Nutter$^{1}$,	
B. Sibthorpe$^{6}$, 
J. Di Francesco$^{7,8}$, 
A. Abergel$^{9}$, 
D. Arzoumanian$^{3}$,  
M. Benedettini$^{10}$,
S. Bontemps$^{11}$,  
D. Elia$^{10}$,
M. Hennemann$^{3}$,
T. Hill$^{3}$,
A. Men'shchikov$^{3}$,
F. Motte$^{3}$, 
Q. Nguyen-Luong$^{3,12}$,
N. Peretto$^{3}$, 
S. Pezzuto$^{10}$,
K. L. J. Rygl$^{10}$, 
S. I. Sadavoy$^{7,8}$,  
E. Schisano$^{10}$, 
N. Schneider$^{11,3}$, 
L. Testi$^{13}$,
G. White$^{14,15}$ }}\vspace{0.1cm}\\
$^{1}$ Cardiff School of Physics and Astronomy, Cardiff University, Queens Buildings, The Parade, Cardiff, Wales, CF24 3AA, UK \\
$^{2}$ Jeremiah Horrocks Institute, University of Central Lancashire, PR1 2HE, UK \\
$^{3}$ Laboratoire AIM, CEA/DSM--CNRS--Universit\'e Paris Diderot, IRFU/Service d'Astrophysique, CEA Saclay, 91191 Gif-sur-Yvette, France \\
$^{4}$ CNRS, IRAP, 9 Av. Colonel Roche, BP 44346, 31028 Toulouse Cedex 4, France \\
$^{5}$ Universit�e de Toulouse, UPS-OMP, IRAP, 31028 Toulouse Cedex 4, France \\
$^{6}$ UK Astronomy Technology Centre, Royal Observatory Edinburgh, Blackford Hill, Edinburgh, Scotland, EH9 3HJ, UK \\
$^{7}$ National Research Council Canada, Herzberg Institute of Astrophysics, 5071 West Saanich Road, Victoria BC Canada, V9E 2E7 \\
$^{8}$ Department of Physics and Astronomy, University of Victoria, PO Box 355, STN CSC, Victoria BC Canada, V8W 3P6 \\
$^{9}$ Institut d'Astrophysique Spatiale, CNRS/Universit\'e Paris-Sud 11, 91405 Orsay, France \\
$^{10}$ Istituto di Astrofisica e Planetologia Spaziali, Via Fosso del Cavaliere 100, 00133 Roma, Italy \\
$^{11}$ Universit\'e de Bordeaux, Laboratoire d'Astrophysique de Bordeaux, CNRS/INSU, UMR 5804, BP 89, 33271, Floirac Cedex, France \\
$^{12}$ Canadian Institute for Theoretical Astrophysics, University of Toronto, 60 St. George Street, Toronto, ON, M5S 3H8, Canada \\
$^{13}$ ESO, Karl-Schwarzschild Str. 2, 85748 Garching, Germany \\
$^{14}$ Space Science and Technology Department, Rutherford Appleton Laboratory, Chilton, Didcot, Oxon OX11 0QX, UK \\
$^{15}$ Department of Physics \& Astronomy, The Open University, Milton Keynes MK7 6AA, UK 
}

\begin{document}
	\maketitle
	\begin{abstract}
		The whole of the Taurus region (a total area of 52 sq. deg.) has been observed by the  {\it Herschel} SPIRE and PACS instruments at wavelengths of 70, 160, 250, 350 and 500~$\mu$m as part of the {\it Herschel} Gould Belt Survey. In this paper we present the first results from the part of the Taurus region that includes the Barnard 18 and L1536 clouds. A new source-finding routine, the Cardiff Source-finding AlgoRithm (\textsc{csar}), is introduced, which is loosely based on \textsc{clumpfind}, but that also generates a structure tree, or dendrogram, which can be used to interpret hierarchical clump structure in a complex region. Sources were extracted from the data using the hierarchical version of CSAR and plotted on a mass-size diagram. We found a hierarchy of objects with sizes in the range 0.024--2.7\,pc. Previous studies showed that gravitationally bound prestellar cores and unbound starless clumps appeared in different places on the mass-size diagram. However, it was unclear whether this was due to a lack of instrumental dynamic range or whether they were actually two distinct populations. The excellent sensitivity of {\it Herschel} shows that our sources fill the gap in the mass-size plane between starless and pre-stellar cores, and gives the first clear supporting observational evidence for the theory that unbound clumps and (gravitationally bound) prestellar cores are all part of the same population, and hence presumably part of the same evolutionary sequence (c.f. Simpson et al. 2011).
	\end{abstract}

\section{Introduction}

The Gould Belt is a large ($\sim1$\,kpc diameter) loop of molecular clouds and OB associations that is inclined at $\sim20\degr$ to the Galactic Plane \citep{1847herschel, 1879gould}. The Gould Belt is important for star  formation studies as it contains most of the nearby (d~$<$~500~pc) low-mass star formation regions, where our observations have their highest spatial resolution and mass sensitivity. Thus the Gould Belt is an excellent laboratory in which to study all stages of low-mass star formation. For these reasons, separate consortia from the {\it Herschel} \citep{2010andre}, {\it Spitzer} \citep[Allen et al. in prep]{2009evans}, and JCMT \citep{2007wtb} communities have mapped, or are mapping, significant fractions of the Gould Belt at wavelengths from 3.6\, \micron\ to 850\, \micron. 

This paper presents the first {\it Herschel} Gould Belt Survey \citep{2010andre} observations of the Taurus Molecular Cloud. This cloud contains $\sim2-4\times10^{4}$\,M$_{\odot}$ of molecular material distributed in a series of filaments and cores \citep{2008goldsmith, 2008kenyon}. 
The observed star formation efficiency in Taurus is $\sim$1\% 
\citep{2008goldsmith}. An analysis of {\it Planck} data towards Taurus found line-of-sight averaged dust temperatures of ${\sim}17$\,K and ${\sim}14$\,K in the diffuse and dense parts respectively \citep{2011A&A...536A..25P}. \citet{2010rebull} used {\it Spitzer} observations to compile a catalogue $\sim$350 new and previously known YSO candidates across the Taurus cloud. The established distance to Taurus is 140\,pc  \citep{1980straizys,1994kenyon,2008loinard}.

\begin{figure}
	\centering{
		\includegraphics[width=\columnwidth]{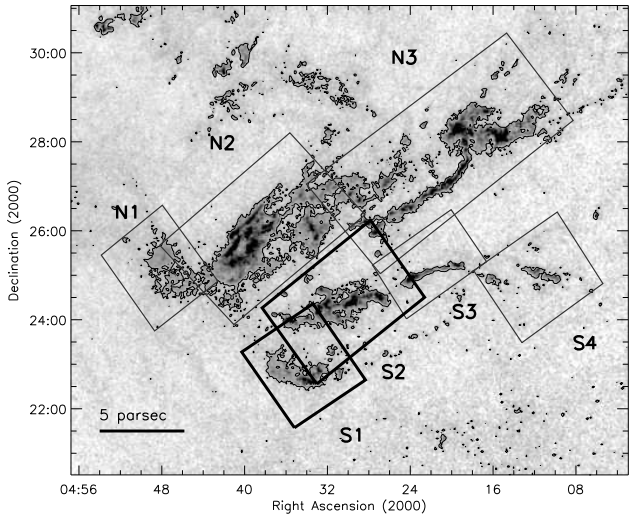}
	}
	\caption{A coverage map of the {\it Herschel} Gould Belt Survey \citep{2010andre} of the Taurus star formation region. The greyscale and contours show visual extinction calculated from the 2MASS survey \citep{2011schneider}. Plotted over this are the foot-prints and names of the {\it Herschel} fields. The regions presented in this paper are outlined by the thick lined boxes. Note that this image covers $\sim10^\circ\times10^\circ$. A single $A_v=3$ contour is plotted. }
	\label{fig1}
\end{figure}

Fig~\ref{fig1} shows an extinction map of Taurus produced from 2MASS data \citep{2011schneider}, overlaid with the fields in our {\it Herschel} survey. The {\it Herschel} fields cover the northern and southern Taurus filaments (labelled N1-N3 and S1-S4 in our survey) and five satellite complexes (L1489, L1517, L1544, L1551, and IRAM 04191; not shown in Fig~\ref{fig1}). The total area is 52 sq. deg. making it the largest region mapped in the Herschel Gould Belt Survey. In this paper, we present data for the first two fields from Taurus to have been observed, reduced, and analysed, namely S1 and S2. These two  fields are outlined in thick black boxes in Fig~\ref{fig1}.

Populations of molecular cores found by different tracers in star forming clouds occupy different regions of the mass-size plane. \citet{2001motte} found that cores seen in the submillimetre continuum and clumps seen by molecular line tracers such as CO lie in very different regions of this plane. There are two possible explanations for this difference: either selection effects are at play, or there are two fundamentally different populations of object, i.e., one of low density clumps that rarely form stars \citep[c.f. ][]{2010wt}, and one of higher density cores that are the precursors to star formation, normally referred to as prestellar cores \citep{1994wsha,2007wta}. In this paper we will use the first look Taurus dataset to test {\it Herschel}'s ability to populate this mass-size plane and examine whether it is possible to demonstrate which of the two posssible explanations is more likely. 

\section{Observations}
\label{observations}

\subsection{Data Reduction}

\begin{figure*}
		\includegraphics[width=\textwidth]{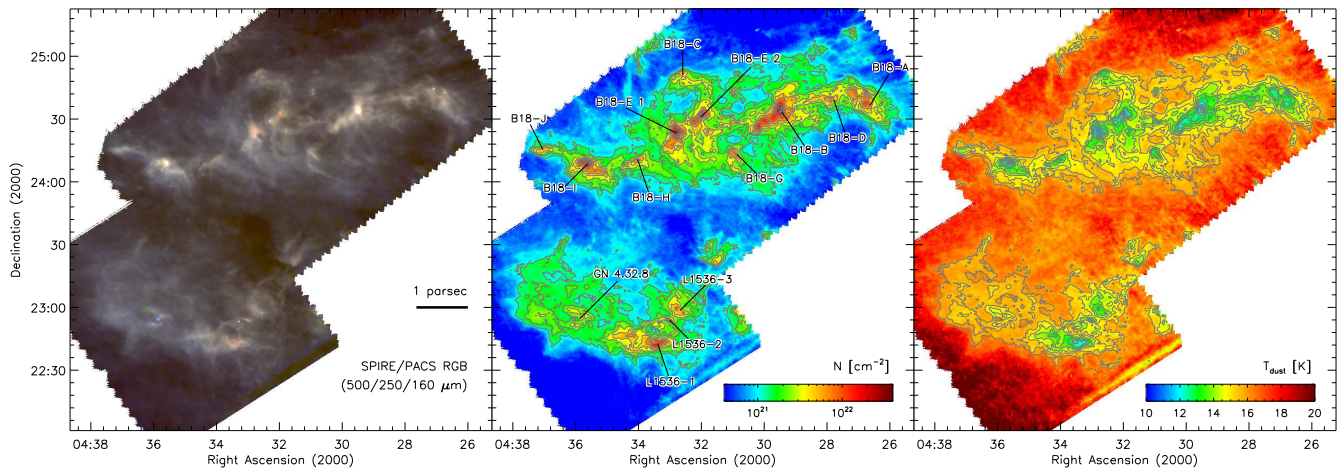}
\caption{Maps of the S1 and S2 regions. {\bf Left:} A false-colour image with red, green and blue showing the 500-$\mu$m, 250-$\mu$m and 160-$\mu$m data respectively. {\bf Middle and Right:} Column density (middle) and colour temperature (right) maps of the S1 and S2 regions derived from the {\it Herschel} 70-$\mu$m to 500-$\mu$m and {\it Spitzer} 24-$\mu$m data (\citealt{2007gudel}; see text for details). The contours show column density starting at 5-$\sigma$ with each subsequent contour level being 1.5 times the previous level. The colour bar indicates the scale. Two large elongated filaments can be seen, along with smaller-scale structure. The northern filament is known as Barnard~18 or TMC~2 \& 3. The southern filament is known as L1536 at its western end and reflection nebula GN~04.32.8 at the eastern end. The bluish point source coincident with the reflection nebula in the left panel is the star HP~Tau. The clumps within B18 are labelled A-J \citep{1988heyer} in the middle panel. The clumps in L1536 are numbered and the reflection nebula is also marked. }
	\label{maps}
\end{figure*}

\begin{table}
	\caption{\label{tab} Multi-wavelength properties of the Herschel maps. The first line lists the native instrumental FWHM. The second line list the median absolute monochromatic intensities towards the S1-S2 fields as derived from IRAS/Planck observations. The third line lists the 1$\sigma$ rms levels as measured from the maps after they have been convolved/resampled to the 500$\mu$m resolution and pixel-grid. }  
\begin{tabular}{ l ccccc }
\hline 
Property & \multicolumn{5}{c}{Wavelength} \\
	& 500$\mu$m & 350$\mu$m & 250$\mu$m & 160$\mu$m & 70$\mu$m  \\
\hline
Beam FWHM (\arcsec) 	& 36 	& 25 	& 18 	& $12\times16$ 	& $6\times12$ 	\\
Median (MJy/sr) 	& 10.7 	& 23.4  & 38.8 	& 60.0 		& 3.97 		\\
1$\sigma$ (MJy/sr)	& 0.8	& 1.5	& 2.2	& 2.9		& 3.8		\\
\hline
\end{tabular}
\end{table}

{\it Herschel} is a space telescope with a 3.5m-diameter mirror that operates in the far-infrared and submillimetre regimes  \citep{2010pilbratt}. The observations for this paper were taken simultaneously with the Photodetector Array Camera and Spectrometer, PACS \citep{2010poglitsch}, and the Spectral and Photometric Imaging Receiver, SPIRE \citep{2010griffin,2010swinyard} using the combined fast-scanning (60 arcsec/s) SPIRE/PACS parallel mode. These data used PACS filters centred at wavelengths of 70-$\mu$m and 160-$\mu$m, with angular resolutions of 6\arcsec$\times$12\arcsec\ and 12\arcsec$\times$16\arcsec\ respectively (accounting for the scan speed), and SPIRE filters centred at wavelengths of 250~$\mu$m, 350~$\mu$m, and 500~$\mu$m, with angular resolutions of 18\arcsec, 25\arcsec\ and 36\arcsec\ respectively. The angular resolutions are also listed on line 1 of Table \ref{tab}. The final PACS and SPIRE angular resolutions are equivalent to spatial resolutions of 0.008--0.024~pc (1,700--5,000~AU) at the assumed distance to Taurus of 140~pc. 

The data were taken on Observing Day 275 (2010 February 13th) and have  {\it Herschel} Observation IDs of 1342190652--55. The total observation duration was 8 hours, 42 minutes. Flicker noise arising from the instrument electronics, so called 1/f noise, introduces a slowly varying component to the detector timelines. This was suppressed by mapping each field in two orthogonal scan directions. The data timelines from each map were reduced separately and then combined at the map-making stage.

The SPIRE data were processed with the software package \textsc{hipe}, version 8.1\,\citep{2010ott} until the `level 1' stage. The SPIRE observation baseline was removed using the `destriper' module in \textsc{hipe} and a correction for the relative gain of each SPIRE bolometer was applied. Maps were reconstructed using the `naive' map-making algorithm. The extended source calibration was used. The rest of the PACS reduction and map reconstruction was done using version 16 of \textsc{scanamorphus} \citep{2012roussel}. These maps were found to be within 5\% of the equivalent reconstruction using the alternative `Madmap' mapmaker \citep{2010ott}.

\subsection{Map morphology}

The left-hand panel of Fig~\ref{maps} shows the {\it Herschel} data of the S1 and S2 region as a false  colour image. The three wavelengths used to make this image are 500~$\mu$m  (shown in red), 250~$\mu$m (shown in green) and 160~$\mu$m (shown in blue) convolved to the 500~$\mu$m resolution. The data show two approximately parallel filamentary structures, with the northern filament (coincident with the S2 field) being significantly longer than the southern filament (coincident with the S1 field). In addition, many smaller filaments and arcs can be seen emanating from the two main filaments. Many warm compact sources, which appear blue, can also be seen within the data. The middle and right-hand panels of Fig~\ref{maps} show the column density and colour temperature maps of the same region derived from the data through smoothing the data to a common 36\arcsec\ resolution and fitting the resulting SED of each common pixel with a greybody ($\beta=2$; see Section \ref{sed} for more details). Once again, the two main filaments can be seen. Objects mentioned in this paper are labelled on the middle panel of Fig~\ref{maps}.

The northern filament has a projected length of $\sim3\degr$ ($\sim$7~pc) and appears roughly linear on the plane of the sky. This filament is known as Barnard~18 \citep[hereafter B18;][]{1919barnard}, and was divided into clumps labelled B18-A to B18-I by \citet{1988heyer} using a map of $^{13}$CO emission; these clumps are indicated on Fig~\ref{maps}. Clump B18-E splits into two on our maps and they are labelled E1 and E2 here. B18-F is a faint feature in the {\it Herschel} data and is not labelled. B18-J is a new feature seen here for the first time and is labelled on Fig~\ref{maps}. The southern filament splits roughly into two halves. The western half is usually known as Lynds Dark Nebula~1536, or L1536 \citep{1962lynds}. The eastern half is known as the reflection nebula GN~04.32.8  \citep{2003magakian}. The morphology of this filament appears to change in Fig~\ref{maps} from L1536 to GN~04.32.8. L1536 appears broadly similar to the clumps in the B18 filament, but in the region of GN~04.32.8 the filament is fainter. 

The reflection nebula GN~04.32.8 contains the star HP~Tau which appears as a point source surrounded by a crescent shaped feature in the Herschel maps, suggestive of interaction between the point source and the cloud. \citet{2009torres} used the VLBA to measure a parallax distance to HP~Tau of 161.2 $\pm$ 0.9 pc. The interaction of the star with the cloud would imply that this part of the cloud is also at that distance. The plane-of-sky separation of the two filaments is $\sim$2$^\circ$ which, at 161~pc, would equate to about 5.5~pc. There is some suggestion in CO surveys, however, that the filaments are not at the same distance \citep{2007stojimirovic,2008narayanan}. Given that uncertainty we use, as stated earlier, the established distance of 140\,pc for the entire cloud \citep{1980straizys,1994kenyon,2008loinard}. If the southern filament is at ${\sim160}$\,pc and the northern filament is closer to the canonical ${\sim140}$\,pc then the two filaments could be separated by up to $\sim$20~pc along the line of sight.

Comparing the left-hand panel of Fig~\ref{maps} with {\it Herschel} observations of some other regions  in the Gould Belt, interesting differences appear. For example, compared to the Aquila region \citep{2010andre}, the region displayed in Fig~\ref{maps} shows relatively little change in colour, and thus temperature, between different parts of the cloud. The coldest, densest parts appear slightly redder than the average, with only isolated points of blue (i.e., warm YSOs). In contrast, the equivalent image of Aquila \citep{2010andre} shows far more variation, with large areas appearing red and the warm W40 nebula showing up clearly in blue. This difference is not an artefact of the false-colour imaging; it represents a real difference between the Aquila and Taurus star-forming regions. There is far more variation in temperature across the Aquila region \citep[$\sim$10--30\,K, partially due to W40; ][]{2010bontemps} than in Taurus ($\sim$10--15\,K).

The mass of the Taurus molecular cloud ($2-4\times10^4\,\mathrm{M}_\odot$ -- \citealt{2008goldsmith, 2008kenyon}) and the Aquila molecular cloud ($4\times10^4\,\mathrm{M}_\odot$ -- \citealt{2010bontemps}) are comparable, although the total area mapped by Herschel in Taurus is ${\sim}40$\% larger. Note that this study only presents a small part of that area. Aquila is also a region with on-going low-mass star formation with a core-mass function that peaks at ${\sim}0.7,\mathrm{M}_\odot$. See Section~\ref{mrsec} for a discussion of the masses of cores in this part of Taurus.

\section{\textsc{Column Density Map}}

\subsection{Pixel-by-pixel SED fitting}
\label{sed}

The column density, $N(\textrm{H}_2)$, map of the Taurus S1-S2 region was reconstructed by fitting the spectral energy distribution (SED) of each pixel in the {\it Herschel} map. Before fitting the pixel SEDs we must calibrate the data in surface brightness units, account for the map zero-point, and then colour-correct each pixel. The default calibration of the PACS data is already in surface brightness units. We converted the SPIRE data from point-source to extended calibration and then to surface brightness units using the recommended factors \citep{2011bendo}. 

Fitting pixel-by-pixel SEDs can be sensitive to systematic offsets between the different wavelengths. For the SPIRE or PACS data alone, such offsets would not be a problem, as a common baseline removal/mapping technique  should remove  a common background. However, different techniques are used for the  SPIRE and PACS instruments. Therefore, it is necessary to perform some measure of large-scale background correction at each wavelength, e.g., by masking a source and estimating a local background \citep[e.g., ][]{2010peretto}, globally filtering the map, or using additional data to fit a zero-point. We chose the latter as it allows us to fix the background over the entire area. We therefore followed the method described by \citet{2010Bernard} to estimate the absolute flux level towards the S1-S2 region from {\it Planck} and {\it IRAS} data.  The median absolute fluxes for each wavelength are listed on line 2 of Table \ref{tab}.

Once the backgrounds had been established, the data were convolved to a common resolution (the SPIRE $500$\,$\mu$m PSF, 36\arcsec\ FWHM) using the \citet{2011aniano} convolution kernels and then co-aligned on the $500$\,$\mu$m pixel-grid (14\arcsec\ pixel width). We colour-correct each pixel using the standard SPIRE \citep{2011bendo} and PACS \citep{2011muller} factors in an iterative cycle. The spectral index at each wavelength is calculated and the required colour correction factor is interpolated from the tabulated values. The correction is then applied to the data and a new set of spectral indices are calculated. This corrected set of indices are in turn used to calculate an updated correction. The spectral-index/correction-factor cycle is repeated until the correction factors converge or a maximum of 20 cycles is reached.

For each position where there are greater than 3\,$\sigma$ detections between 160--500$\mu$m we fit the function 
\begin{equation} 
F_\nu = \frac{M B_\nu(T) \kappa_\nu}{D^2} , 
\end{equation} where $F_\nu$ is the  flux density at frequency $\nu$, $M$ is the mass-per-pixel, $B_\nu(T)$ is the Planck Function for a blackbody with temperature $T$, and $D$ is the distance to the source \citep{1983hildebrand}. The dust mass opacity $\kappa_\nu$ is paramaterized as $\kappa_\nu\propto\nu^\beta$ and is referenced against a value of 0.1\,cm$^2$\,g$^{-1}$ at 1\,THz \citep{1990beckwith}, which assumes a gas-to-dust mass ratio of 100. The distance to Taurus is $D=140$\,pc and we assume that $\beta=2$. There are therefore only two free parameters $T$ and $M$ for each fit. The mass-per-pixel is converted to a column density by,
\begin{equation}
N(\textrm{H}_{2}) = \frac{ M }{m_{\textrm{\small H}} \mu A }  ,
\end{equation}
where $N(\textrm{H}_{2})$ is the column number density of H$_2$, $m_{\textrm{\small H}} \mu$ is the mean particle mass ($m_{\textrm{\small H}}$ is the mass of a hydrogen atom and $\mu$ was taken to be 2.86, assuming the gas is $\sim70\%$ H$_2$ by mass), and $A$ is the area of each pixel. 

The Levenberg-Marquardt least-squares minimisation package MPFIT \citep{2009markwardt} was used for the SED fitting. We use a single temperature fit over the range 160--500$\mu$m. This neglects line-of-sight temperature variations and gives a temperature that is representative of the entire column of material, not just its coldest component, see \citet{2009shetty} for a discussion. We explored a 2-component temperature fit by including MIPS \citep{2004rieke} 24\,$\mu$m data from the  Taurus {\it Spitzer} Legacy Program \citep[made available from the  {\it Spitzer} Science Centre]{2007gudel}. For positions where there are 3\,$\sigma$ detections at 24~$\mu$m, 70~$\mu$m, and 160~$\mu$m, it was possible to fit the SED at 24-500~$\mu$m with the sum of two components. In practise this only applied close to the 70\,$\mu$m compact sources as these fields contain no extended-area of strong short wavelength emission. The cold component fit was virtually identical to the original 160--500$\mu$m fit so we only use a single component fit.

Fig~\ref{maps} shows the $N(\textrm{H}_{2})$ and $T$ results (for the cold component). Column density contours, which start at 5-$\sigma$ and increase geometrically with each increment being 1.5 times the previous contour level, are shown on the middle and right maps for reference. The temperature map shows minima  of $\sim10$\,K in B18-E 2, B18-E 1, and L1536-1, with the rest of the higher-density filamentary material lying at a temperature of $\sim12-14$\,K. Away from the filaments the temperature rises to $\sim16$\,K. The first analysis of {\it Planck} data towards Taurus found a similar temperature towards dense regions \citep{2011A&A...536A..25P}. \citet{2005delBurgo} used IRAS 60\,$\mu$m and  ISOPHOT 200\,$\mu$m maps of the TMC-2 clump to infer two distinct  dust temperature components, a warm 20\,K component which followed the 60\,$\mu$m emission and a colder 12.5\,K dense component that followed the 200\,$\mu$m emission. Their colder component is similar to that which we measure towards the filaments. 

\subsection{Estimation of Uncertainty}

The intensity uncertainty of each pixel at each wavelength is taken as the quadratic sum of three components. The first component is $\sigma_{rms}$, the pixel-to-pixel rms measured on the resampled, regridded intensity map. This was calculated by subtracting a map smoothed by twice the beam FWHM from the original map and measuring the standard deviation of the remaining intensity. The calculated values of $\sigma_{rms}$ are listed in Table \ref{tab}. A map uncertainty product is available from the Herschel map making processes, but it was found that the PACS uncertainty maps contained striping associated with masked/bad pixels. While this effect is small (${\sim}10$\% of the overall uncertainty) it is correlated with the scan direction and, if included in the fitting, propogates to the output column density map. Therefore, we instead use the per pixel rms across all wavelengths. This method avoids any scaling that would have been necessary due to confusion effects from the resampling/regridding process. 

The other two uncertainty components are related to the intensity calibration. As described above, the total intensity is the sum of two components -- the Herschel measured intensity $F_{H}$ at each pixel and the absolute background level of the map $F_{B}$. Each of these has a separate systematic fractional uncertainty. We assume that the emission is extended, so the systematic fractional error $C_H$ is 12\% for the SPIRE bands \citep{2011vaktchanov} and 10\% for the PACS 160$\mu$m \citep[][]{2012paladini}. The fractional error on the absolute background is $C_B = 5$\% \citep{2010Bernard}.  

Therefore, a pixel with total intensity $F_\nu = F_H + F_B$, will have a total uncertainty given by  
\begin{equation}
	\sigma_\nu^2 = \sigma_{rms}^2 + (F_H C_H)^2 + (F_B C_B)^2 	
\end{equation}
In practise the second term dominates in regions of high intensity while the last term dominates in regions of low intensity. Only the first term is used when determining whether the intensity of a pixel is statistically detected or not. The full value of $\sigma_\nu$ is used for the SED fitting. The uncertainty on the $N(\textrm{H}_{2})$ fit is dominated by a systematic component of ${\sim}10$\% above a column density of $10^{21}$\,cm$^{-2}$, a value similar to the systematic error in the intensity. The per pixel rms of the column density, calculated in the same manner as for the original intensity maps, is $0.16\times10^{21}$\,cm$^{-2}$. The uncertainty on the $T$ fit was found to be ${\sim}0.3$\,K in regions where $T<13$\,K and ${\sim}1$\,K in regions were $T>17$\,K. 

We tested our fitting procedure using synthetic SEDs with data-points matching the {\it Herschel} bands.  These were generated for a random-temperature  $T_{\textrm{\small input}}$ in the range log(5/K)--log(100/K). Noise was added to the data-points and the synthetic SED plus noise was used to obtain $T_{\textrm{\small fitted}}$ in the above manner. In each case, the mean difference between the input and fitted temperature  increased with the input temperature as the peak of the SED moved to  shorter wavelengths outside of the {\it Herschel} bands. Indeed, the mean difference increased until a breakdown temperature, above which the synthetic SED could not be recovered. Our typical calibration error is 10-12 per cent plus a component from the background. For the conservative case of 15 per cent noise, we cannot recover temperatures above $\sim25$\,K when using just 250-500\,$\mu$m data-points and $\sim40$\,K when 160\,$\mu$m data-points are added.

\section{Results}

\subsection{Source identification}
\label{csar}

The maps contain many sources that are extended relative to the beam-size at each wavelength. The extraction of extended sources from regions of complex emission is a difficult process, and is one that many {\it Herschel} surveys will face. There has been much debate as to which method is best for identifying extended sources in submillimetre surveys  \citep[e.g., see ][]{2009pineda, 2010reid}. A comprehensive study of the current state of the art in this area has been carried out by the SPIRE Consortium within the {\it Herschel} team, and it is  still ongoing. For this first-look paper, we use a conservative variant of the well-known routine \textsc{CLUMPFIND} \citep{1994williams}, which we have called the Cardiff Source-finding AlgoRithm\footnote{CSAR is written in IDL and can be downloaded from \url{http:\\star.jmkprime.org}}. (\textsc{CSAR}). \textsc{CSAR} also retains information pertaining to the structure tree of sources within the field making it similar to other recent dendrogram \citep{2008rosolowsky} or sub-structure codes \citep{2010perettoB}. Where \textsc{CSAR} differs is that the data are analysed as a pixel stream. This eliminates the need to pre-filter the map maxima and reduces the number of free parameters involved in its operation. 

\begin{figure}
	\includegraphics[width=\columnwidth]{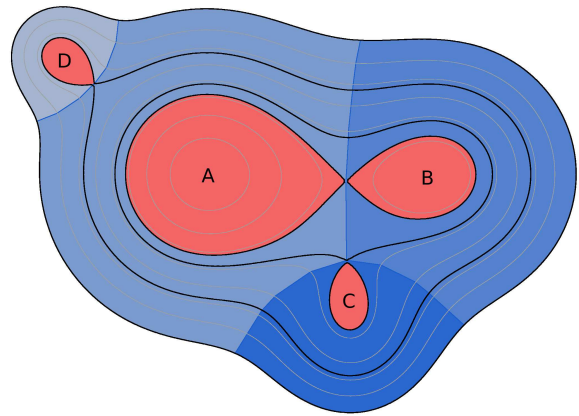} 
	\caption{An illustration of how \textsc{csar} differs from  \textsc{clumpfind} \citep[after fig 2. of ][]{1994williams}.  There are four peaks A-D in this example. The red regions show  the areas around each peak which would be assigned to them by the  non-hierarchical version of \textsc{csar}. The four different blue  shaded regions show the additional area around each peak which would  be assigned to them by \textsc{clumpfind}. The black contours show the clump divisions from the hierarchical version of \textsc{csar}. }
	\label{fig:williams94}
\end{figure}
 
\textsc{csar} is a seeded region-growth \citep[c.f.][]{1994adams} style segmentation algorithm which can be run in either hierarchical or non-hierarchical modes. It is designed to  identify contiguous regions, or clumps, on an intensity map that are  considered significant by virtue of being:  (i) resolved, namely that the clump area is greater than the telescope  beam footprint;  (ii) having no resolved internal sub-structure; and  (iii) having a peak-to-clump minimum contrast ratio greater than 3\,$\sigma$, where  $\sigma$ is the mean pixel-to-pixel error. The clumps that the  non-hierarchical version of $\textsc{csar}$ finds are broadly equivalent  to \textsc{clumpfind}'s seed clumps before the application of its  `friends-of-friends' algorithm \citep{1994williams}. The differences  between \textsc{csar} and \textsc{clumpfind} are illustrated in  Fig~\ref{fig:williams94} which is based on fig~2. of  \citet{1994williams}. The red regions are the clumps found by the non-hierarchical version of $\textsc{csar}$ while the blue regions  show the clumps that would have been found by \textsc{clumpfind}.

The non-hierarchical version of $\textsc{csar}$ works by sorting each  pixel $i$ in the map into a list $S_i$, in order of decreasing intensity.  The first pixel in the list, $S_1$, has the highest intensity on the map  and is assigned a clump index $N=1$. Each pixel in the list is then  considered in turn. The $i$th pixel is assigned a new clump index if it  is a local maximum, i.e., its neighbouring pixels have not already been  assigned a clump index. A pixel that adjoins an existing clump takes that  clump's index, i.e., its neighbouring pixels share a single unique clump  index. 

Clumps grow as the routine proceeds down the intensity gradient, as each new pixel is considered. This process continues until  two clumps collide, i.e., where the  $i$th pixel has immediate neighbours with more than one unique clump  index. This point is equivalent to the position in Fig~\ref{fig:williams94}  where the red regions around peaks A and B touch. The $i$th pixel is then a bridging pixel between two clumps and the code  must make a choice as to which clump it should be assigned to. Each clump is tested  to see if it is significant against the three criteria described above.  If both clumps are significant, then they are marked as finished and are  considered to have reached their maximum size before becoming blended.  They are then excluded from further growth. If one or both of the clumps are not significant, then the clumps are merged. Every pixel is processed  in turn until a preset minimum intensity level, or the minimum on the map  is reached. Each surviving unmerged clump is then checked to make sure it matches the resolution and contrast criteria and any that straddle the edges of the mapped region are excluded. 

The hierarchical version of the \textsc{csar} algorithm uses the unblended  clumps found by the non-hierarchical version as the tips of a structure  tree or dendrogram. \citet{1992houlahan} proposed the use of {\it structure  trees} to measure hierarchical structure within ISM clouds. These were  constructed by decomposing a flux map into discrete grey-level masks.  Clumps were found independently at each grey-level and then cross-matched with overlapping clumps at the N-1 level. \citet{2008rosolowsky}  extended structure trees to work with three-dimensional isosurfaces in  molecular line maps. A true dendrogram, as shown by \citet{2008rosolowsky}, uses an arbitrary x-axis (e.g., a sorted, deprojected clump index) to show  the hierarchical structure of another quantity (temperature, intensity, etc)  on the y-axis \citep{2009goodman}.

\begin{figure*}
	\centering{
	\includegraphics[width=0.7\textwidth]{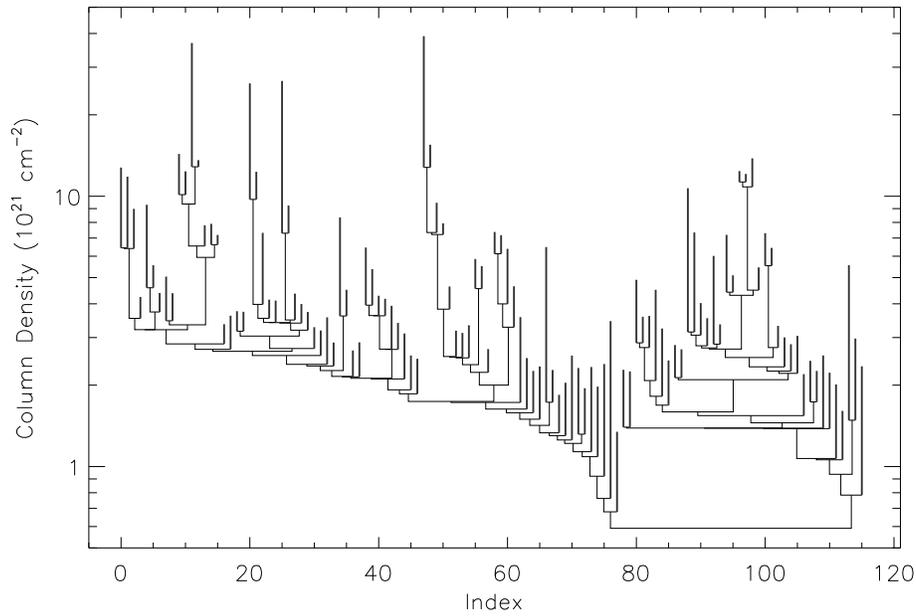}
	}
	\caption{The full dendrogram for the S1-S2 N(H$_{2}$) column density map from  Fig~\ref{maps}. The vertical axis shows column density. Individual sources are shown by the vertical lines, the height of each line shows the range of column densities within that source. Horizontal lines connecting two sources together denote that those sources have merged at that column density. The horizontal axis is an arbitrary clump index which allows the structure tree be plotted as a dendrogram (i.e., a tree diagram without crossed branches). } 
	\label{fig:csar1}
\end{figure*}

\begin{figure*}
	\centering{
	\includegraphics[width=0.6\textwidth]{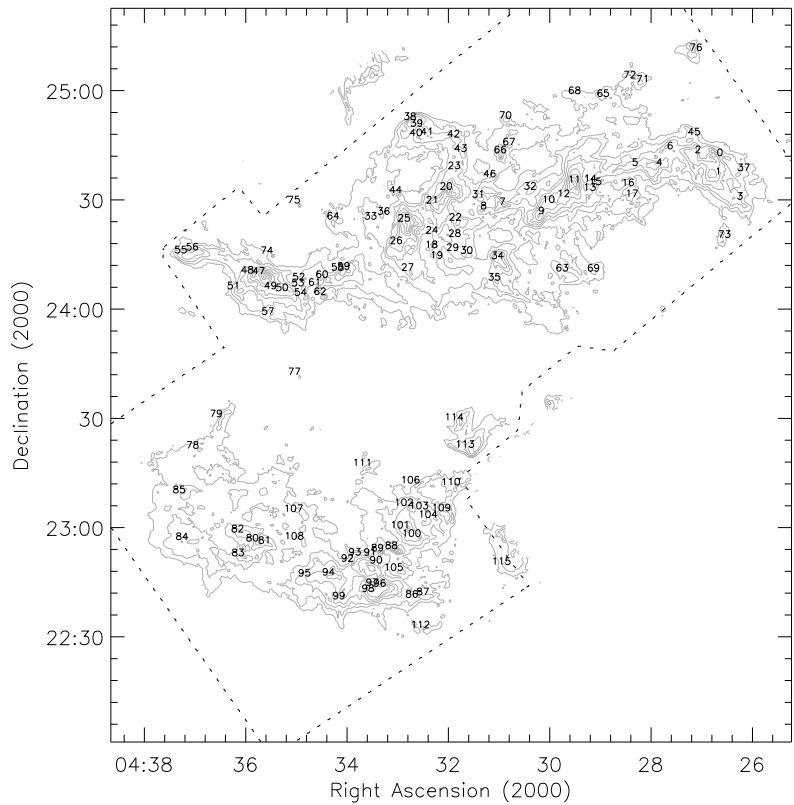}	
	}
	\caption{A re-projection of the clumps as shown in the Figure~\ref{fig:csar1}, but now plotted over the S1-S2 N(H$_{2}$) column density contours from Fig~\ref{maps}. The numbers show the arbitrary index from Figure~\ref{fig:csar1}. It can be seen that, broadly, clumps with indices 1-76 are coincident with the northern filament and cores with indices 78+ are coincident with the southern filament. The dashed contour encloses the features included in the dendrogram. } 
	\label{fig:csar2}
\end{figure*}

\begin{figure*}
	\centering{
		\includegraphics[width=0.8\textwidth]{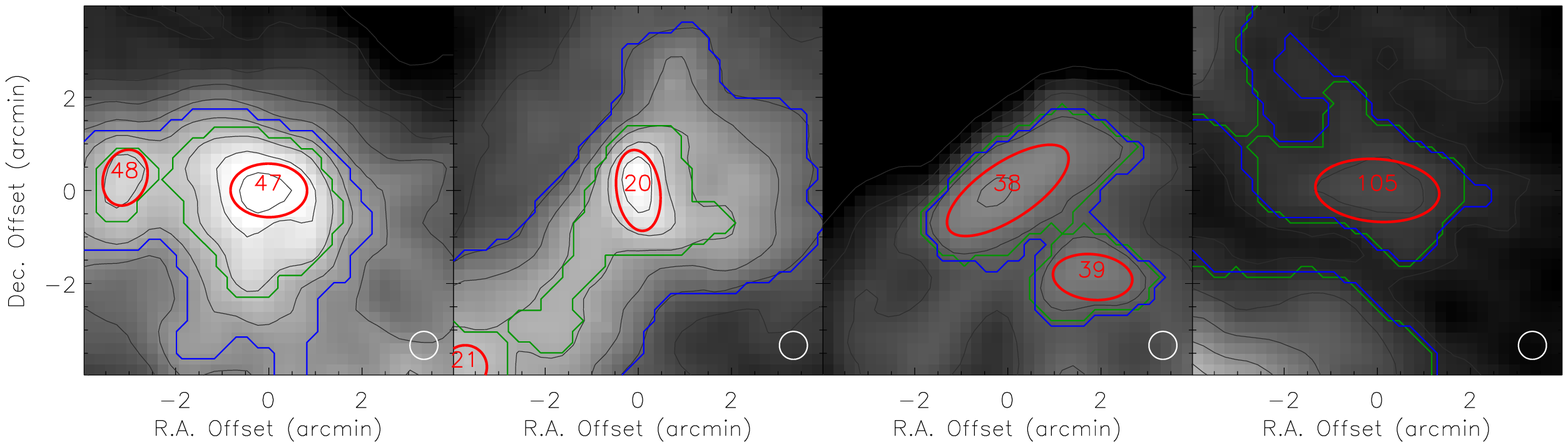}
	}
	\caption{Column density maps for four CSAR cores. The greyscale and contours show $\log_{10}[N(\textrm{H}_{2})/\textrm{cm}^{-2}]$. The greyscale range is 21.1--22.4. The contours are the same as for Fig~\ref{maps}. Each panel shows a source described in the text (centred at 0,0) and its sibling (the source it merges with in the dendrogram). Each core is labelled in red with the same index as shown in Fig~\ref{fig:csar1} and its FWHM contour is shown by the red ellipse. The green-contour is the column density at which these sources merge. This is the same as the horizontal connecting lines on Fig~\ref{fig:csar1}. The blue-contour is the column density at which these merged sources would in turn merge with a third source. The white circle is the telescope beam FWHM. }
	\label{figb3}
\end{figure*}	

It is possible to use the hierarchical version of \textsc{csar} to estimate the shared contribution from the surrounding emission, thereby overcoming any problems of the non-hierarchical version, which makes no attempt to deblend sources or to trace the wings of emission down into the surrounding cirrus.  The assignment of clump indices and merging of non-significant clumps  proceeds as for the non-hierarchical version except that colliding  significant clumps are now allowed to merge. The state of those clumps is recorded at the point of collision. The black contours in  Fig~\ref{fig:williams94} show the clump boundaries at the point of  merger. Moving down the intensity scale, clumps A and B merge at the  topmost black contour, the combined clump containing them then merges  with clump C at the level of second highest black contour. In this  manner a network of merging clumps is traced. Each clump merger is a  node in the structure tree with the unblended non-hierarchical clumps  forming the upper-most branches of the tree.  

\textsc{csar} was run on the column density map of the S1-S2 region shown in Figure~\ref{maps}. The results are shown in Figure~\ref{fig:csar1} and ~\ref{fig:csar2}. Figure~\ref{fig:csar1} shows the dendrogram for those sources while Figure~\ref{fig:csar2} shows a map of the sources extracted. Over-plotted on Figure~\ref{fig:csar2} are column density contours (with levels repeated from Figure~\ref{maps}). 

\textsc{csar} found a single clump hierarchy with two dominant branches (northern and southern filaments), the connection between these is shown by the lowest horizontal-line at $\sim6\times10^{20}$\,cm$^{-2}$ on Figure~\ref{fig:csar1}. The clumps with indices below 77 are part of the B18 branch while clumps with indices above 77 are part of the L1536 branch. For each clump, we compute a position (R.A, Dec) from the mean centre of the clump's half-power contour above its merger column density. The index of each clump is plotted at its position on Figure~\ref{fig:csar2}. The border of each clump is effectively an isocontour and this level is used as that core's background level. The level is calculated from the ring of pixels immediately adjacent to each clump (the need to terminate a clump's growth with an unassigned pixel automatically means that no clumps abut pixels from another clump). The integrated count (either mass or flux density depending on the map type) of each clump is then computed by subtracting the isocontour boundary level from each pixel in the clump and summing the remainder.

The size for all \textsc{csar} sources has been taken as the deconvolved effective radius of a circle with the same area as the source. Fig~\ref{figb3} shows column density maps extracted from Fig~\ref{maps} for a sample four CSAR cores with approximately the same effective radius. The two cores on the left (cores 47 and 20) are coincident with the scatter of blue-triangles in Fig~\ref{mr}. They have masses in excess of 2~$M_{\odot}$ and show evidence for being cooler at their centres (see Fig~\ref{maps}). The two cores on the right (cores 38 and 105) have approximately the same deconvolved size, $\sim0.06$\,pc, as the first two, but have masses less than 2~$M_{\odot}$. 

The two higher-mass cores have SIMBAD entries and are coincident with objects 35 and 31 from the H$^{13}$CO$^{+}$ survey of \citet{2002onishi}. Core 47 has a Herschel mass of 5.4\,$M_\odot$ and radius of 0.06\,pc while the H$^{13}$CO$^{+}$ core is it coincident with has a virial mass of 1.5\,$M_\odot$ and a radius of 0.04\,pc. Virial masses scale with $R^2$ (c.f. \citealt{1981larson}), so scaling the \citet{2002onishi} virial mass for Core 47 to the {\it Herschel} radius gives a mass of 3.4\,$M_\odot$. Core 20 and the H$^{13}$CO$^{+}$ core it is coincident with have the same radius, 0.06\,pc, but core 20 has a mass of 2.1\,$M_\odot$ and the H$^{13}$CO$^{+}$ core has a virial mass of 0.6\,$M_\odot$. Thus, the Herschel mass for each core is higher than the virial mass computed from H$^{13}$CO$^{+}$ as would be expected if both cores were gravitationally bound.

Conversely, the lower-mass core 38 is coincident with object 31 from the CO survey of \citet{1996onishi} and core 105 is not coincident with any extended structure in the SIMBAD database\footnote{Core 105 is coincident with the star 2MASS J04325387+2248375.}. The mass and size of CO object 31 is significantly higher than for the Herschel core (12 versus 1\,\,$M_\odot$ and 0.21 versus 0.058\,pc). Therefore, the CO core appears to be equivalent to an entire sub-branch of the Herschel dendrogram and not a single leaf-node. Figure~\ref{fig:csar1} and \ref{figb3} show that the same {\it Herschel} observations are able to map simultaneously both well-known dense cores and previously uncatalogued lower density cores, and that the structure in the {\it Herschel} maps can be characterised by \textsc{CSAR}.    

\textsc{CSAR} was applied to the column density maps in its hierarchical mode and structures were found on scales in the range 0.024--2.7\,pc. This result is to be expected, since molecular clouds generally contain structure on a variety of scales. A total of 236 nested-sources (which we call nodes) were found in the hierarchy, 115 of which do not contain resolved substructure. A dendrogram of the hierarchy is shown in Figure~\ref{fig:csar1}. 

We also ran the \textsc{getsources} algorithm on our data \citep{2012menshchikov}. This routine found the same sources as \textsc{CSAR}, and others besides. \textsc{getsources} uses all five wavelengths simultaneously to identify sources, and we saw that most of the additional sources it found were seen only in a subset of wavelengths. Furthermore, \textsc{getsources} did not find the large clumps found by \textsc{CSAR} because it does not define sources as such. Any source containing substructure is broken into its components by \textsc{getsources}. Consequently, all of the sources discussed here were found by both algorithms. We believe this is the most robust way to proceed.

\subsection{Mass-Size relation}
\label{mrsec}

\begin{figure*}
	\centering{
		\includegraphics[width=0.9\textwidth]{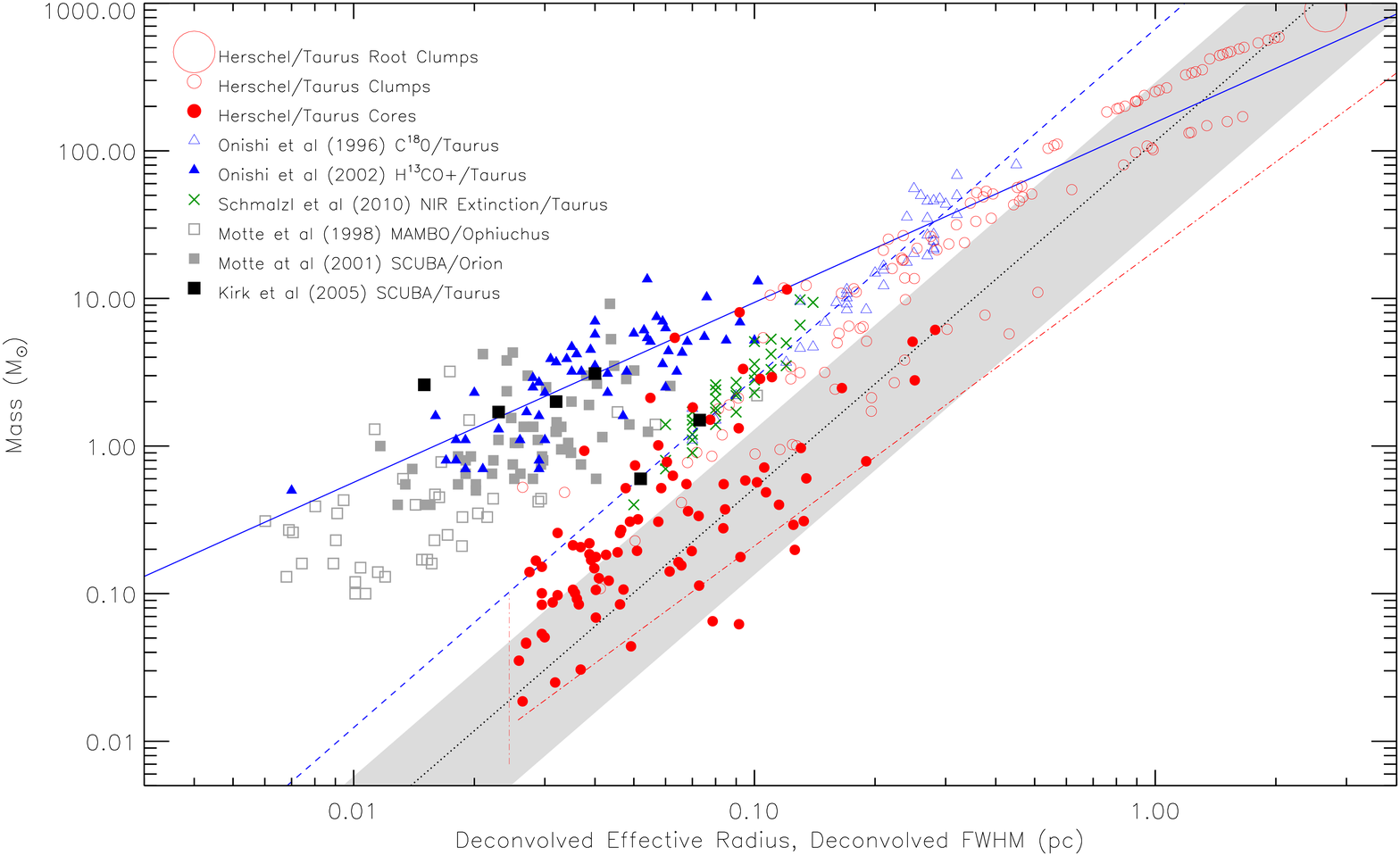}
	}
	\caption{A mass-size plot for resolved sources found in our survey, compared to those found in other surveys using different tracers. The open and closed red circles are the new data from this paper. Comparison data are taken from: \citet{1996onishi}; \citet{1998motte}; \citet{2001motte}; \citet{2002onishi}; \citet{2005kirk}; and \citet{2010schmalzl}. The key at upper-left shows the meaning of each of the symbols. The solid gray band and the dotted line show the $M_\textrm{CO} \propto R_{\textrm{CO}}^{2.35}$ trend for CO cores \citep{1996elmegreen}. The solid and fashed blue lines are fits to the \citet{2002onishi} and \citet{1996onishi} data respectively. The vertical red dot-dashed line is the resolution limit of the Herschel data, the diagonal red dot-dashed line is the mass sensitivity limit for a 3-$\sigma$ Gaussian source with FWHM equal to the deconvolved radius. See text for discussion. It is assumed that the effective radius and FWHM are approximately equal measures of a core's size.}
\label{mr}
\end{figure*}

Fig~\ref{mr} shows the mass-size relation for the sources that we have found \citep[following fig.~4 of][]{2001motte}. The solid red circles indicate the individual resolved objects with no visible substructure and the open red circles denote the objects that we found with substructure. The two types of object have median masses of 0.20, and 25\,M$_\odot$ respectively and median sizes of 0.055 and 0.35\,pc respectively. The root clump -- the total of everything in the hierarchy -- has a mass of 880\,M$_\odot$ and an equivalent size of 2.7\,pc. For comparison, the open and solid grey squares show data for prestellar cores in Ophiuchus \citep{1998motte}  and Orion \citep{2001motte} respectively, observed in the millimetre/submillimetre continuum. Note that the same assumptions regarding mass calculations from flux densities were used in calculating the masses for Orion and Ophiuchus as for the current survey of Taurus. Hence, any offsets between \citet{1998motte}, \citet{2001motte} and the current data are not due to different assumptions about mass opacities, or equivalent systematic uncertainties. 

We use the column density map to calculate masses. The alternative method would be to calculate the core's flux densities at each wavelength using a local background and then fit the SED. We checked what difference this second method would have on the leaf-node sources by fitted SEDs (using the same method as for the pixel-by-pixel SEDs) to background subtracted flux densities measured using the \textsc{csar} extraction contours. It was found that using a local background increased the core masses by ${\sim}$20-30\% on average. This increase is due to the removal of flux associated with warmer, overlaying material, However, using this method discards the information content of the dendrogram, which we wish to preserve, and requires one to have a priori knowledge of the core locations. The advantage of the dendrogram/column density method is that it preserves information on the global properties of the region and supplies local core properties in a single self-consistent operation.   

Analysts of molecular cloud structure have pursued the idea that there exists a power-law relationship between the mass of a core, $M$, and its radius $R$, such that $M\propto R^{k}$. \citet{1981larson} found $n(\textrm{H}_{2})\propto R^{-1.1}$, which is equivalent to $M\propto R^{1.9}$, based on a sample of mainly $^{13}$CO observations. \citet{1996elmegreen} found a value of $k=2.35$ for a sample of clouds, which they studied in CO. A study of CO clouds in the inner-galaxy Molecular Ring found a power-law with $k=2.36\pm0.04$ \citep{2010roman}. However, the power-law  exponent as measured by denser gas tracers (e.g. the submillimetre  continuum) appears to be  different. For example, \cite{2010curtis} found for dense cores in Perseus that a power-law exponent value closer to $\sim$1.5 was a better fit to their data.  \citet{2009pineda} showed that the exponent can be influenced by the choice of source extraction algorithm. 

The dataset of \citet{1996elmegreen} and \citet{1998motte, 2001motte} did not include data from Taurus, so we also include in Fig~\ref{mr} literature data from two other surveys of Taurus.  The open blue triangles show cores in Taurus observed by \citet{1996onishi} in C$^{18}$0 (1--0). These points can be fit by a power-law relationship with an exponent of $k=2.4\pm0.1$, shown as a blue dashed line on Fig~\ref{mr}.  This value of 2.4 is similar those found for CO clouds by other authors \citep{1996elmegreen,2010roman}.

\citet{2002onishi} carried out an H$^{13}$CO$^{+}$ emission survey of Taurus. The higher critical density for collisional excitation of H$^{13}$CO$^{+}$ (1--0), of  $\sim10^{5}$\,cm$^{-3}$, over C$^{18}$O (1--0), of $\sim10^{4}$\,cm$^{-3}$, allowed \citet{2002onishi} to probe denser parts of the same clouds. The filled blue triangles in Fig~\ref{mr} show the virial masses calculated from these data. They also appear to follow a power-law trend. We fit to this trend a power-law relationship with an exponent value of  $k=1.2\pm0.1$ (solid blue line). This value is consistent with that found for dense cores in Perseus by \citet{2010curtis}. The H$^{13}$CO$^{+}$ masses are virial masses ($M \propto R$, assuming weak or no linewidth-size correlation, c.f. fig 7. of \citealt{2002onishi}), so the value of this exponent in not unexpected.

The lowest mass \textsc{CSAR} cores are consistent with the trend for CO clumps while the highest mass \textsc{CSAR} cores are in the parameter regime of the H$^{13}$CO$^{+}$ (1--0) sources. This shows that, for a given radius, we are sensitive to sources over a range of 2 dex in mass. A number of the substructured \textsc{CSAR} clumps are consistent with the $k=2.4$ power-law fitted to the C$^{18}$O sources. This means that we can detect the same material as the C$^{18}$O (1--0) survey, albeit over a wider range of spatial scale. Thus, the {\it Herschel} \textsc{CSAR} sources are coincident with all of the mass-size regimes shown by the other tracers. 

\section{Discussion and Conclusions}

\citet{2008pineda} showed that C$^{18}$O remained unsaturated below $A_v{\sim}10$ and \citet{2010pinedajl} showed that CO and dust masses for parsec sized clumps were in good agreement. We see a similar effect in Fig~\ref{mr} where the clumps from our dendrogram that have radii of  ${\sim}0.3$\,pc have masses which are comparable to CO clumps of the same size \citep{1996onishi}. However, the case is different at smaller scales where interpretation of individual cores is hampered by the spatial filtering of ground-based bolometer observations \citep{2010pinedaje}, the relationship between CO mass and $A_v$ becomes non-linear above $A_v>10$\,mag \citep{2010pinedajl}, and by the differing chemical history of each core \citep{2011hacar}. The high-spatial range of Herschel and the reliability of dust mass measurements makes it the perfect telescope to observe objects on a wide range of scales. 

\citet{2010kauffmann} produced a mass-size diagram using a dendrogram technique for multiple clouds. They concluded that the mass-size data were not consistent with a constant column density relationship ($k=2$) and that cores in low-mass star formation regions had masses that lie below a threshhold with an exponent of $k=1.33$ (close to the solid blue line in Figure \ref{mr}). However, \citet{2010kauffmann} had to use two different data-sets to bridge the same parameter space. What Herschel adds to this type of analysis is its ability to sample a larger range of spatial scales in a single map and to extract a greater dynamic mass range of object for a given radius. Nevertheless, the linear part of the mass-size relation in the top-left of Figure~\ref{mr}, as the structure tree folds into the root node, does appear to be inconsistent with a global value of $k=2$ in agreement with \citet{2010kauffmann}. 

As mentioned previously, \citet{2001motte} found that cores seen in the submillimetre continuum (squares on Fig~\ref{mr}) and clumps seen by molecular line tracers such as CO (grey shaded area on Fig~\ref{mr}) lie in very different areas of the mass-size plane. Our Herschel data allows is to study the two possible explanations for this difference: either selection effects are at play, or there are two fundamentally different populations of object, i.e., low density non-star forming starless cores \citep[c.f. ][]{2010wt}, and higher density prestellar cores \citep{1994wsha,2007wta}.

If the latter explanation were correct, then the populations would remain distinct in the mass-size plane, even after selection effects were corrected. However, {\it Herschel} is sensitive to much lower column density material than  previous ground-based submillimetre continuum arrays. Consequently, the sensitivity limit of {\it Herschel} continuum data lies further to the lower right on Fig~\ref{mr}. Hence, {\it Herschel} effectively removes the old selection effect problems, and so its data  should be able to settle this issue.

Previous results from {\it Herschel} observations of other regions have been moderately inconclusive in resolving this issue. In more active star-forming regions such as Aquila \citep{2010andre} and Vela \citep{2012giannini}, the cores found by {\it Herschel} lie in the same part of the diagram as the ground-based submillimetre continuum cores, indicating that they are mostly prestellar cores. In non star-forming regions, such as the Polaris, the cores found by {\it Herschel} lie in the same band of the diagram as the CO clumps \citep{2010andre}, indicating that they were mostly CO clumps \citep[c.f. ][]{2010wt}. A gap remained in the mass-size relation plot between these two samples.

Now we can see from Fig~\ref{mr} that the {\it Herschel} cores in the Taurus star-forming region lie largely in the gap between the ground-based continuum cores and CO clumps. Their appearance rules out the possibility that there are two distinct populations of cores. Instead, the {\it Herschel} data indicate that all cores are part of the same  population. Furthermore, this conclusion leads to the hypothesis that lower density cores can evolve into prestellar cores by migrating across this diagram before collapsing to form protostars. A suggested form of this evolution has been previously proposed by \citet{2011simpson}. Further results from the {\it Herschel} Gould Belt Survey in other regions should help to resolve the exact manner of the  evolution of prestellar cores.

\section*{Acknowledgements}

SPIRE has been developed by a consortium of institutes led by Cardiff Univ. (UK) and including: Univ. Lethbridge (Canada); NAOC (China); CEA, LAM (France); IFSI, Univ. Padua (Italy); IAC (Spain); Stockholm Observatory (Sweden); Imperial College London, RAL, UCL-MSSL, UKATC, Univ. Sussex (UK); and Caltech, JPL, NHSC, Univ. Colorado (USA). This development has been supported by national funding agencies: CSA (Canada); NAOC (China); CEA, CNES, CNRS (France); ASI (Italy); MCINN (Spain); SNSB (Sweden); STFC, UKSA (UK); and NASA (USA).
PACS has been developed by a consortium of institutes led by MPE (Germany) and including UVIE (Austria); KU Leuven, CSL, IMEC (Belgium); CEA, LAM (France); MPIA (Germany); INAF-IFSI/OAA/OAP/OAT, LENS, SISSA (Italy); IAC (Spain). This development has been supported by the funding agencies BMVIT (Austria), ESA-PRODEX (Belgium), CEA/CNES (France), DLR (Germany), ASI/INAF (Italy), and CICYT/MCYT (Spain).
This work is based [in part] on observations made with the Spitzer Space Telescope, which is operated by the Jet Propulsion Laboratory, California Institute of Technology under a contract with NASA.
J.M.K. also wishes to acknowledge STFC for post-doctoral funding support while this research was carried out.
D.E., K.L.J.R., and E.S. are funded by an ASI fellowship under contract numbers I/005/11/0 and I/038/08/0.

\end{document}